\documentclass[a4paper,preprint,aps,superscriptaddress]{revtex4}

\usepackage{graphicx}% Include figure files
\usepackage{dcolumn}% Align table columns on decimal point
\usepackage{bm}% bold math\setlength{\itemsep}{0pt}

\newcommand{\eqa}{\begin{equation}}
\newcommand{\eqz}{\end{equation}}
\newcommand{\eqma}{\begin{eqnarray}}
\newcommand{\eqmz}{\end{eqnarray}}

\begin{document}
\newcommand{\e}{{\em e}~}
\title{Basis set convergence of post-CCSD contributions to molecular atomization energies}
\author{Amir Karton}
\affiliation{Department of Organic Chemistry, 
Weizmann Institute of Science, 
IL-76100 Re\d{h}ovot, Israel. Email: {\tt comartin@weizmann.ac.il}}
\author{Peter R. Taylor}
\affiliation{Department of Chemistry and Centre for Scientific Computing,
University of Warwick,
Coventry CV4 7AL, United Kingdom} 
\author{Jan M. L. Martin*}
\affiliation{Department of Organic Chemistry, 
Weizmann Institute of Science, 
IL-76100 Re\d{h}ovot, Israel.  Email: {\tt comartin@weizmann.ac.il}}

\date{Received May 9, 2007; accepted June 12, 2007;
{\em J. Chem. Phys.}, in press: AIP MS ID {\bf 310728JCP}}

\begin{abstract}
Basis set convergence of correlation effects on molecular atomization energies beyond the CCSD (coupled cluster with singles and doubles) approximation has been
studied near the one-particle basis set limit. Quasiperturbative
connected triple excitations, (T), converge more rapidly than $L^{-3}$ 
(where $L$ is the highest angular momentum represented in the basis set), while higher-order connected triples, $T_3-(T)$, converge
more slowly --- empirically, $\propto L^{-5/2}$. Quasiperturbative
connected quadruple excitations, (Q), converge smoothly as $\propto
L^{-3}$ starting with the cc-pVTZ basis set, while the cc-pVDZ basis set causes
overshooting of the contribution in highly polar systems. Higher-order
connected 
quadruples display only weak, but somewhat erratic, basis set
dependence. Connected quintuple excitations converge very rapidly with
the basis set, to the point where even an unpolarized double-zeta
basis set yields useful numbers. In cases where fully iterative
CCSDTQ5 (coupled cluster up to connected quintuples) calculations are not an option, CCSDTQ(5) (i.e., coupled cluster up to connected quadruples plus a quasiperturbative connected quintuples correction) cannot be relied upon in the 
presence of significant nondynamical correlation, whereas 
CCSDTQ(5)$_\Lambda$ represents a viable alternative. Connected quadruples
corrections to the core-valence contribution are thermochemically
significant in some systems. We propose an additional variant of W4 theory [A. Karton, E. Rabinovich, J. M. L. Martin, and B. Ruscic, {\em J. Chem. Phys.}
{\bf 125}, 144108 (2006)], denoted W4.4 theory, which is shown to yield an RMS deviation from
experimental atomization energies (active thermochemical tables, ATcT) of only 0.05 kcal/mol for systems for which ATcT values are available. We conclude that ``$3\sigma\leq 1$~kJ/mol'' thermochemistry is feasible with current technology, but that the more ambitious goal of $\pm$10~cm$^{-1}$ accuracy is illusory, at least for atomization energies.
\end{abstract}

\maketitle

\section{Introduction}

There exists an extensive literature on one-particle basis set
convergence at the SCF (self-consistent field) and CCSD (coupled cluster with all singles and doubles) levels. Basis set convergence at the
SCF level is fairly rapid (except for `inner polarization' issues
caused by back-bonding into $d$ orbitals of 2nd row elements in high
oxidation states, see\cite{cl2o7tae} and references therein), and at
least for atoms and diatomic molecules, exact numerical solutions are
available on a semi-routine basis\cite{JensenSCF}.

Basis set convergence of the MP2 and CCSD correlation energies is
likewise well studied: the main reference point here is explicitly
correlated 
quantum chemistry, which exhibits 
vastly more rapid basis set
convergence than standard one-particle Gaussian basis sets (see e.g.,
Valeev\cite{valeev} for a very recent review).  While some authors
argue in favor of supplanting Gaussian basis sets altogether with
explicitly correlated methods, others have shown that very high
accuracy can be achieved by judicious combination of very large
Gaussian basis sets with extrapolation techniques that are motivated
either by the physics of pair correlation energies in helium-like
systems\cite{Sch63,Hil85,Kut92,Hal98,Klopper} or
empirically\cite{l4,Schwenke2005} (see also Feller and
Peterson\cite{Feller2007} and references therein).

Basis set convergence beyond the CCSD level has been much less well
studied, and remains an issue even for advocates of explicitly
correlated methods, as the computationally efficient extension of the
latter beyond CCSD is a nontrivial challenge. An early paper by
Klopper and co-workers\cite{Hel97} pointed out that at the CCSD(T) level --- i.e., CCSD plus a quasiperturbative triples correction\cite{Rag89,Wat93}, often cited as `the gold standard in quantum chemistry'\cite{goldstandard} --- the (T) term of
the correlation energy converges much more rapidly with the
basis set than the CCSD term. More recent studies that focus at least
partly on the (T) term include the work of Schwenke\cite{Schwenke2005}
and of Crawford et al.\cite{Crawford2006}. Allen and coworkers, in a
string of studies based on their focal-point
approximation\cite{Allen}, have studied convergence in many systems on
an {\em ad hoc} basis. Martin and coworkers\cite{w4} address basis set
convergence for connected quadruple and quintuple excitations in some
detail, although not as close to the basis set limit as is perhaps
desirable (see also Ref.\cite{Feller2007}).

What is missing from the literature at present is a study where for a
number of representative systems convergence for the main post-CCSD
correlation contributions to molecular atomization energies
is considered as close to the one-particle
basis set limit as possible, converged at the level of 0.01~kcal/mol
where feasible.

The present paper reports such a study. It will also serve to provide
additional theoretical support for the approximations inherent in the
W$n$\cite{w1,w1eval,w1cioslowski,w3,w4} and HEAT\cite{heat,heat2}
families of computational thermochemistry protocols. Finally, the
present study should also shed some light on the intrinsic limits to
accuracy with present-day wave function-based {\em ab initio} techniques ---
even if we were to assume, for the sake of argument, that CCSD basis convergence is a solved problem.

\section{Computational methods}

Most calculations reported in the present work were carried out on the
Linux cluster of the Martin group, which consists of machines
custom-built by Access Technologies of Re\d{h}ovot, Israel.  We relied
very heavily on four machines in particular. All have 2 terabytes of
high-bandwidth scratch disk space (eight 250 GB Serial-ATA disks
striped 8-way on a hardware RAID controller). Two of the machines have
quadruple dual-core AMD Opteron 870 CPUs, the remaining two have dual
quad-core Intel Cloverton CPUs at 2.66 GHz. One of these latter
machines has 32 GB of RAM, the remaining three have 16 GB.  Some calculations
were carried out at the University of Warwick, using Opteron-based
systems.

The CCSD(T)  calculations with the aug-cc-pV7Z basis
set\cite{av7z,carbon-av7z} --- which contains up to $k$ functions --- were carried
out using both PSI 3.3.0\cite{psi3} at Weizmann, and a locally
modified version of DALTON 2.0\cite{dalton} at Warwick.  CCSD(T)
calculations in smaller basis sets were carried out using MOLPRO
2006.1\cite{molpro} for closed-shell cases, and the
Austin-Mainz-Budapest version of ACES II\cite{aces2de} for open-shell
cases. All post-CCSD(T) calculations were carried out using an OpenMP
parallel version of K\'allay's general coupled cluster code
MRCC\cite{mrcc} interfaced to the Austin-Mainz-Budapest version of the
ACES II\cite{aces2de} program system.

Unless specifically noted otherwise, unrestricted Hartree-Fock
references were used for open-shell systems, and CCSD(T)/cc-pV(Q+d)Z
reference geometries were taken from Ref.\cite{w4}.

All basis sets employed, except for the unpolarized Dunning-Hay double zeta  (DZ) basis set\cite{DZ} employed for some post-CCSDTQ contributions, belong to the correlation consistent family of Dunning and
coworkers\cite{Dun89,Ken92,Wilson,pwCVnZ}.

The following basis set extrapolations were considered: (a) the simple
two-point $A+B/L^3$ expression of Halkier et al.\cite{Hal98}, which is
rooted in the partial-wave expansion of singlet-coupled pair energies in helium-like
atoms\cite{Sch63,Hil85,Kut92} and is used extensively in both the
W$n$\cite{w1,w3,w4} and HEAT\cite{heat,heat2} families of
computational thermochemistry protocols; (b) Schwenke's empirical
two-point extrapolation formulas\cite{Schwenke2005}, which are
equivalent to $A+B/L^\alpha$ with an empirical extrapolation exponent
$\alpha$; (c) three-point linear extrapolation formulas of the type
$A+B/L^3+C/L^4$ and $A+B/L^3+C/L^5$, similar to those first proposed
in Ref.\cite{l4}. (We also considered the variable-exponent three-point formula $A+B/L^C$, not as an actual extrapolation --- as it is not size-consistent --- but use the `effective decay exponent' obtained as a probe for effective convergence rate, similar to Ref.\cite{AllenConvergence}.)

\section{Results and discussion}

\subsection{Quasiperturbative triple excitations, (T)}

Extrapolated
contributions of ``parenthetical'' triples to the
total atomization energy are given in Table~\ref{tab:parenT}. In a
number of cases, we were able to reach as far as aug-cc-pV7Z basis
sets (AV7Z for short).

In the following discussion, the notation AV\{L-1,L\}Z, for instance,
will indicate $A+B/L^\alpha$ ($\alpha=3$) extrapolation from
aug-cc-pV($L-1$)Z and aug-cc-pV$L$Z basis sets, unless specifically
indicated otherwise. PV\{L-1,L\}Z stands for the same extrapolation,
but from regular cc-pV$L-1$Z and cc-pV$L$Z basis sets.

Comparison of AV\{5,6\}Z and AV\{6,7\}Z data reveals that, with the
exception of singlet C$_2$ (0.007 kcal/mol), the extrapolated
contributions are converged to better than 0.005 kcal/mol.

The extrapolated AV\{T,Q\}Z data are in surprisingly good agreement
with our best limits. (This extrapolation is used for the (T)
contribution in W2, W3, and W3.2 theory, as well as in HEAT345.) It is
perhaps not coincidental (see below) that Schwenke's extrapolation
formula for AV\{T,Q\}Z basis sets is equivalent to an inverse power
extrapolation with exponent $\alpha$=2.99882, which is only
semantically different from $\alpha$=3.

In contrast, the AV\{Q,5\}Z expression used in W4, W4.2, and W4.3
theory tends to slightly overestimate the basis set limit
contribution, by amounts ranging from 0.05 kcal/mol in C$_2$ via 0.03
kcal/mol in N$_2$, B$_2$, and CO and 0.02 kcal/mol in O$_2$, F$_2$,
and H$_2$O to less than 0.01 kcal/mol in HF. Comparison of AV\{Q,5\}Z,
AV\{5,6\}Z, and AV\{6,7\}Z limits suggests that starting with AVQZ
basis sets, $\alpha$=3 extrapolation approaches the basis set limit
from above (in absolute value), i.e., that convergence is faster than $\alpha$=3. This behavior was previously noted by Crawford et al.\cite{Crawford2006}. (For the AV\{5,6,7\}Z basis sets and the atoms \{C,N,O,F\}, we find
effective decay exponents in the 3.57--3.70 range.) 

Schwenke's extrapolation for the (T) contribution was derived from
fitting to best (T) limits for seven systems: Ne, N$_2$, CH$_2$,
H$_2$O, CO, HF, and F$_2$. These were themselves obtained from what he terms
Òf-limitÓ basis sets (saturated to 5 microhartree in each
angular momentum) going all the way up to $i$ functions. His
AV\{Q,5\}Z extrapolation is equivalent to $\alpha=3.60183$. This
definitely remedies the overshooting problem: in systems like H$_2$O
and C$_2$, SchwenkeAV\{Q,5\}Z basically gets the basis set limit spot-on,
while it tends to be slightly low for other systems. SchwenkeAV\{5,6\}Z is
equivalent to $\alpha=3.22788$, and agrees with the available AV\{6,7\}Z limit data to within 0.003 kcal/mol RMS (root-mean-square), compared to 0.009 kcal/mol for SchwenkeAV\{Q,5\}Z and 0.006 kcal/mol (0.004 excluding S$_2$) for the regular AV\{5,6\}Z extrapolation.

Finally, we considered a three-point linear extrapolation
$A+B/L^3+C/L^4$. AV\{Q,5,6\}Z too seems to behave well, albeit with a tendency
to slightly undershoot the available AV\{6,7\}Z limits. AV\{5,6,7\}Z and AV\{6,7\}Z agree to within 0.007 kcal/mol RMS (0.004 kcal/mol excluding S$_2$).

We conclude that the regular $\alpha=3$ extrapolation is appropriate
for AV\{T,Q\}Z and probably AV\{5,6\}Z basis set pairs, but that
Schwenke's extrapolation (equivalent to $\alpha=3.60183$) is more
appropriate for the AV\{Q,5\}Z pair. For the AV\{5,6\}Z pair,
Schwenke's expression (equivalent here to $\alpha=3.22788$) appears to
be as reliable as $\alpha=3$ or may be slightly more so --- the
difference is too close to call.

We also would like to stress that Schwenke's exponents are themselves
the result of a fit, and that effective exponents for his seven
individual species (reverse-engineered for the present work) reveal a
considerable spread. 
In our opinion, obtaining the (T) contribution converged to 0.01 kcal/mol
using a two-point extrapolation from $spdfgh$ and
$spdfghi$ basis sets appears to be feasible.

\subsection{Higher-order connected triple excitations}

Extrapolated 
contributions of higher-order connected triples,
$T_3-(T)$, to the total atomization energy are given in
Table~\ref{tab:fullT}.

We have PV\{5,6\}Z data available for a limited number of
systems. Comparison with their PV\{Q,5\}Z counterparts 
reveals convergence to better than 0.01 kcal/mol, and
suggests than the PV\{Q,5\}Z numbers are very close to the basis set
limit.

Even from PV\{D,T\}Z basis sets (as used in W4 and W4.2 theory),
useful estimates can apparently be obtained, with the notable
exceptions of B$_2$ and, to a lesser extent, singlet C$_2$.

The PV\{T,Q\}Z numbers, however, reveal that convergence in this basis
set size regime is actually {\em slower} than $\alpha=3$. (The
PV\{T,Q\}Z basis set pair is used for this contribution in the HEAT
approach as well as in W4.3 theory.) Fitting against either the
PV\{Q,5\}Z or the available PV\{5,6\}Z limits suggests an effective
$\alpha=2.5$. On purely empirical grounds, we recommend this for extrapolation of the $T_3-(T)$
term from the PV\{T,Q\}Z basis set pair.

\subsection{Parenthetical connected quadruple excitations}

Raw and extrapolated contributions of parenthetical quadruples to the
total atomization energy --- as obtained using the CCSDT(Q) method as
defined in Ref.\cite{Bomble2005} and implemented in
Ref.\cite{KallayGauss} --- are given in Table~\ref{tab:parenQ}.

In highly polar systems like H$_2$O, HF, OH, and BF, the cc-pVDZ basis
set appears to overshoot the contribution: even in such cases, basis
set convergence for (Q) is however monotonic from cc-pVTZ onwards. In other
systems, convergence is monotonic from cc-pVDZ onwards.

Our best available data are PV\{Q,5\}Z extrapolations. Comparison of
PV\{T,Q\}Z and PV\{Q,5\}Z data reveals that they agree very well with
each other, the largest discrepancies being 0.015 kcal/mol for P$_2$ and Cl$_2$, followed by 0.01 kcal/mol for C$_2$ and
0.007 kcal/mol for BN.
This in turn suggests that basis set convergence, from cc-pVTZ
onwards, is well described by the singlet partial-wave formula
$A+B/L^3$. In contrast, PV\{D,T\}Z extrapolations fare poorly (as
previously reported\cite{w3}), the cc-pVDZ basis set being simply too
anemic. The inadequacy of cc-pVDZ is not limited to overshooting in the 
highly polar systems, but extends to severe undershooting in the second-row molecules. 

The $A+B/L^3$ convergence we observe for the (Q) contribution is not
obvious and deserves some attention.  Our (perhaps naive)
rationalization is based on our analysis of the size of the (Q)
contribution in different systems: our results clearly demonstrate
that (Q) is largest in systems with strong nondynamical correlation.
In fact, the systems we have studied here that fall into this category
all have considerable multiconfigurational character and would ideally
be described using several reference configurations when trying to
recover dynamical correlation.  The additional reference
configurations would be double excitations from the nominal
Hartree-Fock configuration, and describing dynamical correlation would
entail double excitations from these additional reference
configurations, or quadruple excitations from Hartree-Fock.  These
systems will thus have large (Q) contributions, but since these are
predominantly double excitations from other reference configurations
we can expect the typical basis set convergence for double
excitations, that is, $\propto L^{-3}$.\cite{PRTfootnote}

As was shown previously\cite{w4}, the cc-pVTZ numbers multiplied by an
empirical scaling factor of 1.1 (as used in W4 theory\cite{w4}) agree
quite well with the basis set limit estimates available. Could one
come up with a solution that is more reliable than cc-pVDZ yet less
costly than scaled cc-pVTZ?  It was noted before\cite{w3} that a
[4s3p1d] Widmark-Malmqvist-Roos\cite{WMR} atomic natural orbital\cite{ANO} basis set appears to be
devoid of the overshooting problems associated with cc-pVDZ. As this
basis set is still considerably smaller than cc-pVTZ, it might offer a
cost-effective alternative, at least for first-row systems. (For second-row systems, ANO431 suffers from the same undershooting defects as cc-pVDZ.)

Finally, we note that brute-force convergence to 0.1 kcal/mol requires
at least cc-pVQZ basis sets, and that brute-force convergence to 0.01
kcal/mol will probably require at least a cc-pV6Z basis set.

\subsection{Higher-order connected quadruple excitations}

It was suggested before\cite{w4}, based on data up to cc-pVTZ, that
higher-order connected quadruple excitations, $T_4-(Q)$, converge
rapidly with the basis set. In the present work, we were able to go
out to cc-pVQZ for a number of species. Results are summarized in
Table~\ref{tab:fullQ}.

It can be seen there that variation between cc-pVQZ, cc-pVTZ, and
scaled cc-pVDZ amounts to a few hundredths of a kcal/mol at most, even for
such pathologically multireference systems as singlet C$_2$\cite{C2} and 
singlet BN\cite{BN}. No clear
way of extrapolating or correcting these data can be seen, and it
should be noted that even the O$_2$ and S$_2$ CCSDTQ/cc-pVQZ calculations
strained our available computational resources to the very limit.

The $T_4-(Q)$ contribution uniformly reduces the atomization energy, and its absolute magnitude is roughly proportional to the degree of nondynamical correlation, varying from essentially nil in cases like HF and H$_2$O via about 0.1 kcal/mol for systems like CO, O$_2$, F$_2$, and P$_2$ to over 1 kcal/mol for the singlet states of C$_2$ and BN. One would expect a contribution that primarily expresses nondynamical correlation effects to exhibit weak basis set dependence --- as we indeed observe.

We considered still further reduction of the basis set to a simple unpolarized double-zeta (DZ) set. Performance then becomes very uneven, however, and the same holds for the cc-pVDZ basis set with the polarization functions removed.

\subsection{Connected quadruples considered as a whole}

Let us now consider all of $T_4$ together. Results are summarized in
the upper pane of Table~\ref{tab:allQ}.

It can be seen here that achieving convergence to within a few
hundredths of a kcal/mol is quite feasible, but that anything beyond
that will be a very arduous task.

The W4.3 combo --- PV\{T,Q\}Z for (Q), PVTZ for $T_4-(Q)$ --- is
generally within 0.01--0.03 kcal/mol of the best achievable basis set
limits. It tends to slightly underestimate in cases like HF and
H$_2$O, but slightly overestimate otherwise (particularly for strongly 
multireference cases like B$_2$, C$_2$, and BN).

The W4 combo\cite{w4} --- PVTZ for (Q), PVDZ for $T_4-(Q)$, both scaled by 1.1
--- overall sacrifices fairly little accuracy for drastic cost
savings. The most problematic first-row system appears to be B$_2$, 
for which an overestimate by 0.08 kcal/mol is seen. Our limited second-row data include some significant differences (0.07 kcal/mol for P$_2$,
0.10 kcal/mol for S$_2$, and 0.08 kcal/mol for Cl$_2$), and illustrate why it is desirable, where feasible, to `walk the extra mile' for W4.3 calculations on second-row systems.

In HEAT345(Q)\cite{heat2} and W4lite\cite{w4}, higher-order quadruples are neglected entirely,
and parenthetical quadruples approximated by a simple CCSDT(Q)/cc-pVDZ
calculation. This works better than it has any right to, in fact, but
significant errors arise for highly multireference systems as well as
those for which the bonding is highly polar, and for second-row compounds. 
The latter two issues reflect 
the limitations of the cc-pVDZ basis set. As for the former issue,
Stanton and coworkers have argued\cite{Bomble2005,heat2,Sta97} that the
CCSDT(Q) method should in fact benefit from an error cancellation
between higher-order quadruples contributions and the complete neglect
of quintuple excitations. This comparison has been made in the lower
pane of Table~\ref{tab:allQ}. We see there that this error
cancellation holds rather well in some cases like C$_2$, but much less
so in cases like B$_2$. Substituting the ANO431 basis set improves
agreement for the highly polar systems. 
It has been shown elsewhere\cite{BeBAlSi} that the HEAT345(Q)/W4lite type
approximation can also lead to very significant errors (up to 0.5
kcal/mol for CS) in second-row systems, and we found here
that substituting ANO431 affords no succor for those
either. Quite simply put, cc-pVDZ is too limited a basis set to
universally and reliably capture quadruple excitation effects.

\subsection{Connected quintuples}

The limiting basis set dependence of CCSDTQ5 calculations is
$O(n^5N^{7})$ (where $n$ is the number of electrons and $N$ the number of basis functions), 
and therefore extended basis set CCSDTQ5 calculations quickly become
intractable. Fortunately, as seen in Table~\ref{tab:T5}, such effects
converge {\em very} rapidly with the basis set --- even a simple DZ basis set
captures the effect to within a few hundredths of a kcal/mol in all
cases. (This again makes sense if the $T_5$ effects are primarily seen as an expression of nondynamical correlation. Results with the cc-pVDZ basis set with polarization functions removed are nearly identical --- as noted previously\cite{w4} --- and afford some additional cost savings, especially in second-row compounds.)

In only five cases were we able to go out to cc-pVTZ --- HF,
B$_2$, C$_2$ ($X~^1\Sigma^+$), BN ($a~^1\Sigma^+$), and N$_2$ --- and in this latter case, the calculation
was only barely feasible on the available hardware. For BN and C$_2$, the
PVDZ-PVTZ differences are 0.03 and 0.02 kcal/mol, respectively; for the remaining systems they are 0.01 kcal/mol or less.

Predictably, the only systems for which one finds chemically
significant connected quintuples contributions are those with
appreciable nondynamical correlation. 

In contrast to the case of $T_4$ --- where CCSDT(Q) is exceedingly
useful --- parenthetical quintuples, CCSDTQ(5),\cite{KallayGauss} are of very limited
utility. They may severely overestimate the effects of $T_5$ in cases
with 
substantial nondynamical correlation, and the CCSDTQ5$-$CCSDTQ(5)
difference still exhibits appreciable basis set dependence in cases
like C$_2$. While additivity approximations like
[CCSDTQ(5)-CCSDTQ]/PVDZ + [CCSDTQ5-CCSDTQ(5)]/DZ seem to work
reasonably well in other cases, their reliability seems far from
assured.

The CCSDTQ(5)$_\Lambda$ method\cite{KallayGauss}, on the other hand,
seems to do a much better job, and is a realistic option in cases
where full CCSDTQ5 calculations would entail
unrealistic CPU time and/or memory requirements. In a recent W4 study
on a number of perfluoro and perchloro compounds\cite{BeBAlSi},
CCSDTQ(5)$_\Lambda$/DZ was employed for the $T_5$ term in BF$_3$, as a
full CCSDTQ5 calculation would have required iteratively solving for
about five billion amplitudes.

Can the calculation of connected quintuples be avoided entirely?
Feller and Peterson\cite{Feller2007} suggest estimating the
contributions beyond CCSDTQ by means of Goodson's continued fraction
expression\cite{Goodson}. We attempted both this and a simple
geometric extrapolation, $E_{\rm FCI}-E_{\rm CCSDTQ} \approx
-\Delta E_Q^2/(\Delta E_Q-\Delta E_T)$, where $E_{\rm FCI}$ denotes
the full CI 
energy. Both expressions have similar (limited) predictive
power: sometimes (e.g., C$_2$) they predict $T_5$ contributions
surprisingly well, sometimes (e.g., F$_2$) they overestimate them by
half an order of magnitude. We also considered both expressions for
the contribution of connected {\em sextuple} excitations, $T_6$, and
there we found both expressions to be of similar quality as explicit
CCSDTQ5(6)/DZ or CCSDTQ56/DZ calculations.

\subsection{Parenthetical triples in core-valence correlation}

The contribution of parenthetical triples to the core-valence
correlation energy may be small in absolute terms, but it is
chemically quite significant in relative terms (molecule vs. separate
atoms) --- and indeed, it has been shown in the past\cite{w1} that as
much as half of the core-valence contribution in total atomization
energies can derive from parenthetical triples.

Basis set convergence for this contribution is summarized in
Table~\ref{tab:coreT}. As can be seen there, this contribution is
nearly saturated at the ACV\{T,Q\}Z level (as used in the W4 family),
and the distance from the basis set limit is on the order of 0.01
kcal/mol 
or less.

\subsection{Higher-order correlation effects in core-valence correlation}

In W4.2 and W4.3 theory, a correction for higher-order triples in the
core-valence contribution is obtained at the CCSDT/cc-pwCVTZ level. In
Table~\ref{tab:coreHO}, we consider both further basis set expansion
for this contribution and the effect of connected quadruples.

First, we compare the core-valence CCSDT-CCSD(T) difference between
CV\{T,Q\}Z and unextrapolated CVTZ. Differences range from essentially
nil for systems dominated by dynamical correlation to as much as 0.1
kcal/mol for pathologically multireference systems like C$_2$ and
BN. The contributions almost universally {\em increase} the total
atomization energy, and tend to roughly cancel with the negative
post-W4.3 correlation contributions in the {\em valence} component.

Secondly, we consider connected quadruples, even if only at the
CCSDT(Q)/CVTZ level. This contribution becomes significant for two
categories of molecules: (a) pathologically multireference systems
like B$_2$ (0.07 kcal/mol), BN (0.12 kcal/mol), and C$_2$ (0.08
kcal/mol); (b) some second-row molecules like Cl$_2$ (0.04 kcal/mol), S$_2$,
and CS (0.08 kcal/mol each). This contribution, too, almost universally
increases molecular binding (PH$_3$ being the only real exception).

\subsection{General observations and W4.4 theory}

In the preceding discussions we have focussed in detail on the many 
individual contributions.  We now step back and take a broader view.

First, many of the post-W4.3 correlation contributions are in
the 0.1 kJ/mol (0.024 kcal/mol) or above range, and their explicit
calculation is simply too arduous a task because of the fierce CPU
time scalings involved. As such, the prospects for `brute force'
calculation of atomization energies
to 10 cm$^{-1}$ seem quite bleak ---
even discounting such issues as small errors in the zero-point vibrational energy (see, e.g., Ref.\cite{c2h6tae} for an illustration), higher-order Born-Oppenheimer corrections, and higher-order relativistic corrections. 

Second, and fortunately, a fair degree of mutual cancellation exists
between the valence correlation improvements on one hand and
inner-shell higher order triples on the other.

This being said, we here incorporate some of our findings in a new
post-W4 method, to be known by the name W4.4 theory. Relative to W4.3
theory defined and discussed in Ref.\cite{w4}, the changes are the
following:
\begin{itemize}
\item Either (variant a) the valence (T) contribution is extrapolated
from AV\{5,6\}Z basis sets, or (variant b) Schwenke's extrapolation
formulas are used for both the singlet-and triplet coupled CCSD pairs
(effective exponents for AV\{5,6\}Z basis sets: $\alpha_S=3.06967$ and
$\alpha_T=4.62528$) as well as for the valence (T) contribution, with
AV\{Q,5\}Z basis sets (effective exponent 3.60183, see above).
\item The $T_3-(T)$ term is extrapolated using $A+B/L^{2.5}$,
  following our observations above;
\item A connected quadruples core-valence term is computed at the
CCSDT(Q)/cc-pwCVTZ level;
\item As it was found to be significant in Ref.\cite{c2h6tae} for systems
with many hydrogen atoms, we add a correlation contribution to
the diagonal Born-Oppenheimer
correction\cite{DBOC,ValeevSherrill}.  We compute this
at the CISD/cc-pVDZ level, which was shown in
Ref.\cite{c2h6tae} to be sufficient for the differential correlation contribution.
\end{itemize}

Results are compared with earlier W4 variants and the best available
ATcT (active thermochemical tables\cite{Branko1,Branko2,Branko3})
values in Table~\ref{tab:w4.4}. The ATcT values themselves were previously
published in Ref.\cite{w4}.

On average, improvements compared to W4.3 are modest. In many cases,
both methods have small errors on opposite sides, with W4.3 being
slightly higher than the ATcT reference value and W4.4 slightly
lower. W4.3 did, however, exhibit large discrepancies of obscure origin from
ATcT for a few systems, such as C$_2$H$_2$ (+0.17 kcal/mol),
N$_2$ (+0.13 kcal/mol), and Cl$_2$ (-0.10 kcal/mol). In W4.4
theory, the discrepancies for C$_2$H$_2$ and N$_2$ are cut by more
than half, while Cl$_2$ stays in place thanks to a compensation
between improving the valence triples (which decreases the binding
energy, and this increases the discrepancy with experiment) and the
inclusion of core-valence quadruples (which significantly increases
the binding energy in this molecule with so many subvalence
electrons). For the systems given in Table~\ref{tab:w4.4}, the RMS
deviation from the ATcT values drops from 0.08 kcal/mol for W4 via
0.07 kcal/mol for W4.3 to 0.05 kcal/mol for W4.4 (both variants). The
latter number implies a 95\% confidence interval of just 0.1 kcal/mol.

There is very little to choose between the two W4.4 variants. The
extra cost of the CCSD(T)/AV6Z calculation in variant (a) could be an
argument in favor of variant (b), but especially for 2nd-row systems,
the extra cost will be dwarfed by that of the core-valence (Q)
calculation. Over the systems surveyed, variant (a) has a slightly
larger maximum positive error than (b) (for C$_2$H$_2$), but a
slightly smaller maximum negative error (for Cl$_2$).

The size of the differences being considered here begs the question
whether errors caused by imperfections in the reference geometry could
not be of a similar magnitude. W4 theory specifies a
CCSD(T)/cc-pV(Q+d)Z reference geometry, which should be well enough
converged for the {\em valence} correlation contribution to the
geometry. However, it has been known for some
time\cite{hf,cc,pwCVnZ,BakGEOM} that inner-shell correlation makes
contributions to typical bond distances on the order of several
milliangstroms, and that all-electron CCSD(T) with the core-valence
weighted cc-pwCVQZ basis set\cite{pwCVnZ} (or the older Martin-Taylor
core correlation basis set\cite{hf}) typically yields bond distances
within about a milliangstrom of experiment. We have recalculated the
total atomization energies for the molecules in Table~\ref{tab:w4.4}
from CCSD(T)/cc-pwCVQZ reference geometries. Essentially all of the
change is confined to the valence and inner-shell CCSD(T) components:
the higher-order correlation terms are barely affected. The
dissociation energies for Cl$_2$ and SO are found to go up by 0.03
kcal/mol, those of C$_2$H$_2$, CO, and N$_2$ by 0.02 kcal/mol, and the
remaining ones by 0.01 kcal/mol or less. 
For some additional species, we found: CO$_2$ 0.03 kcal/mol, CS and
S$_2$ 0.04 kcal/mol, P$_2$ 0.05 kcal/mol. The RMSD for the W4.4b data
at the CCSD(T)/cc-pwCVQZ reference geometries is indeed slightly
reduced, but the difference is not very significant statistically over
this rather small sample.  
(We note that the mean {\em signed} error changes from -0.012 to
+0.003 kcal/mol, i.e., to basically zero.) 
The results suggest that, especially for second-row molecules or
systems with several multiple bonds, the use of CCSD(T)/cc-pwCVQZ
reference geometries may eliminate one potential source of small
errors. For instance, in a recent benchmark study on
P$_4$\cite{P4tae}, we found that the use of a core-valence correlated
reference geometry increases TAE$_0$ by 0.13 kcal/mol. 

Another possible contribution that bears examining at this level of
accuracy is second-order spin-orbit coupling.  For the heaviest system
in our set (Cl$_2$) this was calculated using a multiconfigurational
linear response treatment\cite{Vah92} as implemented in
Dalton\cite{dalton} and found to 
influence the atomization energy by considerably less than
0.01~kcal/mol. 

An independent check is afforded by considering the scaling with the
atomic number $Z$ of the second-order spin-orbit contribution. For the
rare-gas dimers Xe$_2$ and Rn$_2$,  
Runeberg and Pyykk\"o calculated 
second-order spin-orbit contributions to $D_0$ of
+0.7 and +4.5 meV, respectively, while de Jong and coworkers\cite{Dixon2003}
reported contributions of +0.4 and +2.0 kcal/mol, respectively, for
Br$_2$ and I$_2$, and of +0.1 and +0.5 kcal/mol, respectively, for HBr
and HI. These observations suggest approximate $\propto Z^4$ scaling,
which in turn suggests a second-order spin-orbit contribution to
$D_0$(Cl$_2$) of +0.02 kcal/mol.  
Its inclusion would actually improve agreement with experiment
slightly for this system.

\section{Conclusions and perspective}

Basis set convergence of post-CCSD correlation effects has been
studied near the one-particle basis set limit. Quasiperturbative
connected triple excitations, (T), converge more rapidly than $L^{-3}$, 
while higher-order connected triples, $T_3-(T)$, converge
more slowly --- empirically, $\propto L^{-5/2}$. Quasiperturbative
connected quadruple excitations, (Q), converge smoothly as $\propto
L^{-3}$ starting with the cc-pVTZ basis set, while cc-pVDZ causes
overshooting in highly polar first-row systems, and undershooting
in second-row compounds. Higher-order connected
quadruples display only weak, but somewhat erratic, basis set
dependence. Connected quintuple excitations converge very rapidly with
the basis set, to the point where even an unpolarized double-zeta
basis set yields useful numbers. In cases where fully iterative
CCSDTQ5 calculations are not an option, CCSDTQ(5)$_\Lambda$ represents
a viable alternative, while CCSDTQ(5) cannot be relied upon in the
presence of significant nondynamical correlation. Connected quadruples
corrections to the core-valence contribution are thermochemically
significant in some systems. We propose an additional W4 variant,
named W4.4 theory, which is shown to yield an RMS deviation from
experiment (active thermochemical tables, ATcT) of only 0.05 kcal/mol
for systems for which ATcT values are available.

Finally, is it possible to use current technology, brute force, to
calculate molecular atomization energies at the 10~cm$^{-1}$ level?  
Our findings
suggest that the only realistic answer to this question is ``no''.
However, the
more modest goal of ``$3\sigma\leq 1$ kJ/mol'' seems to be not
only realistic, but eminently achievable with methods of the W4
family.

\acknowledgments

Research at Weizmann was funded by the Israel Science Foundation (grant
709/05), the Minerva Foundation (Munich, Germany), and the Helen and
Martin Kimmel Center for Molecular Design. Research at Warwick was 
supported by the Wolfson Foundation
through the Royal Society. JMLM is the incumbent of
the Baroness Thatcher Professorial Chair of Chemistry and a member
{\em ad personam} of the Lise Meitner-Minerva Center for Computational
Quantum Chemistry.  PRT is Royal Society Wolfson Professor of
Chemistry.

The authors would like to thank Prof. John F. Stanton (U. of Texas,
Austin) for helpful discussions and critical reading of an early draft, 
Dr. Branko Ruscic (Argonne National Laboratory) 
for continued encouragement, and Dr. Mih\'aly Kallay (Budapest) for early access to a new version of MRCC and
kind assistance with the code. 

The research presented in this paper is part of ongoing work
in the framework of a Task Group of the 
International Union 
of Pure and Applied Chemistry (IUPAC) on
'Selected free radicals and critical intermediates: thermodynamic
properties from theory and experiment'
(2000-013-2-100, renewal 2003-024-1-100).
See Ref.\cite{iupac1} for further details.

\clearpage

\squeezetable
\begin{table}
\caption{Convergence of the contribution of valence quasiperturbative connected triples, CCSD(T)$-$CCSD, to the total atomization energy (kcal/mol)\label{tab:parenT}}
\begin{tabular}{l|cccccccc}
\hline\hline
  &\multicolumn{4}{c}{A+B/L$^3$}& \multicolumn{2}{c}{Schwenke} & \multicolumn{2}{c}{A+B/L$^3$+C/L$^4$}\\
  & AV$\{$T,Q$\}$Z & AV$\{$Q,5$\}$Z & AV$\{$5,6$\}$Z & AV$\{$6,7$\}$Z & AV$\{$Q,5$\}$Z & AV$\{$5,6$\}$Z & AV$\{$Q,5,6$\}$Z & AV$\{$5,6,7$\}$Z \\
\hline
B$_2$  & 9.809  & 9.794  & 9.768  & N/A    & 9.764  & 9.762  & 9.753  & ---\\
C$_2$  & 19.507 & 19.507 & 19.467 & 19.460 & 19.460 & 19.458 & 19.444 & 19.453\\
N$_2$  & 9.509  & 9.548  & 9.523  & 9.519  & 9.512  & 9.516  & 9.508  & 9.513\\
O$_2$  & 8.381  & 8.414  & 8.394  & 8.391  & 8.373  & 8.386  & 8.380  & 8.387 \\
F$_2$  & 7.688  & 7.700  & 7.685  & 7.681  & 7.666  & 7.678  & 7.673  & 7.677 \\
CO     & 8.120  & 8.145  & 8.122  & 8.118  & 8.115  & 8.116  & 8.108  & 8.114 \\
CN     & 9.687  & 9.720  & 9.700  & N/A    & 9.681  & 9.692  & 9.686  & ---\\
HF     & 2.203  & 2.185  & 2.179  & 2.178  & 2.175  & 2.177  & 2.175  & 2.177 \\
H$_2$O & 3.608  & 3.584  & 3.570  & 3.569  & 3.567  & 3.566  & 3.561  & 3.569 \\
S$_2$&7.166 & 7.254 & 7.228 & 7.215 & 7.210 & 7.219 & 7.207 &  7.200\\
\hline\hline
\end{tabular}
\begin{flushleft}
Unaugmented cc-pV$n$Z basis sets used throughout on hydrogen.\\
C$_2$ and CO AV7Z data obtained using revised AV7Z basis set for carbon\cite{carbon-av7z}.\\
Schwenke AV\{T,Q\}Z numbers are not given explicitly, as they are indistinguishable from the AV\{T,Q\}Z column.\\
aug-cc-pV(7+d)Z basis set for sulfur obtained by expanding even-tempered $d$ series from aug-cc-pV7Z inward with one additional $d$.\\ 
\end{flushleft}
\end{table}
\clearpage

\clearpage
\squeezetable
\begin{table}
\caption{Convergence of the contribution of valence higher-order triples, CCSDT$-$CCSD(T), to the total atomization energy (kcal/mol)\label{tab:fullT}}
\begin{tabular}{l|ccccc}
\hline\hline
 $\hat{T_3}-$(T) & PV$\{$D,T$\}$Z & PV$\{$T,Q$\}$Z & PV$\{$Q,5$\}$Z & PV$\{$5,6$\}$Z & PV$\{$T,Q$\}$Z \\
  & & & & &$\alpha$=2.5 \\
\hline
B$_2$  & 0.240 & 0.113 & 0.079 & 0.088 & 0.080 \\
C$_2$  &-2.194 &-2.248 &-2.287 &-2.291 &-2.292 \\
N$_2$  &-0.778 &-0.756 &-0.773 &-0.778 &-0.779 \\
O$_2$  &-0.543 &-0.497 &-0.526 & N/A   &-0.511 \\
F$_2$  &-0.358 &-0.314 &-0.335 &-0.339 &-0.325 \\
CO     &-0.561 &-0.567 &-0.583 & N/A   &-0.591 \\
CN     & 0.846 & 0.786 & 0.749 & N/A   & 0.760 \\
NO     &-0.355 &-0.335 &-0.354 & N/A   &-0.356 \\
HF     &-0.136 &-0.160 &-0.167 &-0.165 &-0.169 \\
H$_2$O &-0.204 &-0.233 &-0.246 & N/A   &-0.246 \\
P$_2$  &-0.997 &-0.931 &-0.944 & N/A   &-0.957 \\
S$_2$  &-0.498 &-0.482 &-0.484 & N/A   &-0.504 \\
Cl$_2$ &-0.412 &-0.436 &-0.430 & N/A   &-0.456 \\
CS     &-0.635 &-0.636 &-0.645 & N/A   &-0.664 \\
SO     &-0.459 &-0.442 &-0.446 & N/A   &-0.461 \\
ClF    &-0.322 &-0.314 &-0.315 & N/A   &-0.327 \\
\hline\hline
\end{tabular}
\begin{flushleft}
\end{flushleft}
\end{table}
\clearpage

\clearpage
\squeezetable
\begin{table}
\caption{Convergence of the contribution of valence quasiperturbative connected triples, CCSDT(Q)$-$CCSDT, to the total atomization energy (kcal/mol)\label{tab:parenQ}}
\begin{tabular}{l|cccccccc}
\hline\hline
  & PVDZ & PVTZ & PVQZ & PV5Z & PV$\{$D,T$\}$Z &PV$\{$T,Q$\}$Z & PV$\{$Q,5$\}$Z & ANO431 \\
\hline
B$_2$  &0.908 &1.163 &1.220 &1.239 &1.27 &1.262 &1.260 &0.945\\
C$_2$$^a$&2.655 &3.198 &3.311 &3.346 &3.46 & 3.393 &3.382 &2.823\\
BN$^b$ &2.478 &3.041 &3.188 &3.238 &3.28 & 3.296 &3.289 &2.757\\
N$_2$  &1.057 &1.134 &1.217 &1.247 &1.17 & 1.278 &1.279 &1.042\\
O$_2$  &1.122 &1.093 &1.157 &1.179 &1.08 & 1.204 &1.202 &1.040\\
F$_2$  &0.929 &0.912 &0.982 &1.006 &0.91 & 1.033 &1.032 &0.867\\
CO     &0.634 &0.652 &0.700 &0.715 &0.66 & 0.735 &0.731 &0.582\\
CN     &1.237 &1.438 &1.519 &1.544 &1.52 & 1.578 &1.571 &1.249\\
NO     &0.878 &0.913 &0.981 &1.004 &0.93 & 1.031 &1.027 &0.845\\
HF     &0.216 &0.119 &0.132 &0.139 &0.08 & 0.141 &0.145 &0.132\\
H$_2$O &0.261 &0.191 &0.213 &0.223 &0.16 & 0.229 &0.234 &0.213\\
OH     &0.114 &0.078 &0.088 &0.093 &0.06 & 0.095 &0.099 &0.100\\
BF     &0.301 &0.264 &0.290 &0.297 &0.25 & 0.309 &0.304 &0.254\\
CS     &0.590 &0.978 &1.082 &1.119 &1.14 & 1.158 &1.159 &0.472\\
P$_2$  &1.040 &1.431 &1.567 &1.608 &1.60 & 1.666 &1.651 &1.071\\
S$_2$  &0.499 &0.796 &0.899 &0.939 &0.92 & 0.975 &0.980 &0.536\\
Cl$_2$ &0.262 &0.425 &0.487 &0.515 &0.49 & 0.532 &0.545 &0.296\\
\hline\hline
\end{tabular}
\begin{flushleft}
$^a$ $a~^1\Sigma^+_g$ state at $r$=1.24 \AA.\\
$^b$ $X~^1\Sigma^+$ state at CCSDT/cc-pVQZ bond distance, 1.2769 \AA, from Ref.\cite{Watts}.
\end{flushleft}
\end{table}
\clearpage

\clearpage
\squeezetable
\begin{table}
\caption{Convergence of the contribution of valence higher-order quadruples, CCSDTQ$-$CCSDT(Q), to the total atomization energy (kcal/mol)\label{tab:fullQ}}
\begin{tabular}{l|ccccccc}
\hline\hline
 $\hat{T_4}-$(Q) &      &     & W4, W4.2 & W4.3 & best\\
  &PVDZ(no d)&DZ & 1.1$\times$PVDZ & PVTZ & PVQZ \\
\hline
B$_2$  & 0.193 & 0.200 & 0.093 & 0.031 & 0.009 \\
C$_2$  &-1.297 &-1.340 &-1.173 &-1.102 &-1.128 \\
BN$^a$ &-0.827 &-0.828 &-1.226 &-1.187 &-1.214 \\
N$_2$  &-0.177 &-0.191 &-0.171 &-0.151 &-0.166 \\
O$_2$  &-0.088 &-0.056 &-0.137 &-0.128 &-0.146 \\
F$_2$  &-0.084 &-0.058 &-0.116 &-0.113 & N/A \\
CO     &-0.065 &-0.044 &-0.110 &-0.095 &-0.098 \\
CN     &-0.096 &-0.026 &-0.416 &-0.443 &-0.469 \\
HF     &-0.013 &-0.004 &-0.017 &-0.016 &-0.014 \\
H$_2$O &-0.019 &-0.011 &-0.027 &-0.022 &-0.022 \\
OH     & 0.005 & 0.009 & 0.000 &-0.006 &-0.006 \\
P$_2$  &-0.146 &-0.143 &-0.118 &-0.146&-0.169\\
S$_2$  & 0.037 & 0.037 &-0.054 &-0.060 &-0.076\\
Cl$_2$ & 0.007 & 0.007 &-0.025 &-0.020 & N/A \\
\hline\hline
\end{tabular}
\begin{flushleft}
$^a$ At CCSDT/cc-pVQZ bond distance, 1.2769 \AA, from Ref.\cite{Watts}.
\end{flushleft}
\end{table}
\clearpage

\clearpage
\squeezetable
\begin{table}
\caption{All of connected quadruples, $\hat{T_4}$, considered together; connected quadruples and quintuples, $\hat{T_4}+\hat{T_5}$, considered together (all units kcal/mol)\label{tab:allQ}}
\begin{flushleft}
\begin{tabular}{l|cccccc}
\hline\hline
 $\hat{T_4}$ total & W4lite,&  W4,     & W4.3           & better        & best\\
                       &HEAT(Q) &   W4.2    &                &               &     \\
\hline
(Q)                    &PVDZ     &1.1$\times$PVTZ & PV$\{$T,Q$\}$Z &PV$\{$Q,5$\}$Z &PV$\{$Q,5$\}$Z \\
$\hat{T_4}-$(Q)        &null     &1.1$\times$PVDZ & PVTZ           & PVTZ          & PVQZ \\
 \hline
B$_2$  & 0.908  & 1.372 & 1.293 & 1.291 & 1.269 \\
C$_2$  & 2.655  & 2.369 & 2.346 & 2.335 & 2.309\\
BN $^a$& 2.478  & 2.119 & 2.109 & 2.102 & 2.075\\
N$_2$  & 1.028  & 1.027 & 1.056 & 1.049 & 1.034\\
O$_2$  & 1.122  & 1.066 & 1.076 & 1.074 & 1.056\\
F$_2$  & 0.929  & 0.887 & 0.920 & 0.920 & N/A \\
CO     & 0.634  & 0.608 & 0.641 & 0.636 & 0.633\\
CN     & 1.237  & 1.166 & 1.135 & 1.129 & 1.103\\
HF     & 0.190  & 0.104 & 0.112 & 0.115 & 0.117\\
H$_2$O & 0.261  & 0.184 & 0.206 & 0.212 & 0.191\\
P$_2$  & 1.040  & 1.456 & 1.520 & 1.505 & 1.482\\
S$_2$  & 0.499  & 0.822 & 0.915 & 0.920 & 0.904\\
Cl$_2$ & 0.262  & 0.443 & 0.512 & 0.525 & N/A \\
\end{tabular}
\begin{tabular}{l|cccccc}
\hline
 $\hat{T_4}+\hat{T_5}$ & W4lite,&        & W4,     & W4.3           & better        & best\\
                       &HEAT(Q) &        & W4.2    &                &               &     \\
\hline
(Q)                    &PVDZ    & ANO431 &1.1$\times$PVTZ & PV$\{$T,Q$\}$Z &PV$\{$Q,5$\}$Z &PV$\{$Q,5$\}$Z \\
$\hat{T_4}-$(Q)        &null    & null   &1.1$\times$PVDZ & PVTZ           & PVTZ          & PVQZ \\
$\hat{T_5}$            &null    & null   & DZ      & PVDZ           & PVDZ          & PVTZ \\
 \hline
B$_2$  & 0.908 & 0.95 & 1.456 & 1.368 & 1.366 & 1.335 \\
C$_2$  & 2.655 & 2.82 & 2.643 & 2.666 & 2.655 & 2.647\\
BN$^a$ & 2.478 & 2.76 & 2.297 & 2.263 & 2.256 & 2.256\\
N$_2$  & 1.028 & 1.04 & 1.135 & 1.170 & 1.163 & 1.143\\
O$_2$  & 1.122 & 1.04 & 1.142 & 1.179 & 1.177 & \\
F$_2$  & 0.929 & 0.87 & 0.919 & 0.960 & 0.960 & \\
CO     & 0.634 & 0.58 & 0.654 & 0.673 & 0.668 & \\
CN     & 1.237 & 1.29 & 1.293 & 1.253 & 1.247 & \\
HF     & 0.190 & 0.13 & 0.114 & 0.114 & 0.117 & 0.123\\
H$_2$O & 0.261 & 0.21 & 0.190 & 0.214 & 0.220 & \\
P$_2$  & 1.040  & 1.07 & 1.555 & 1.646 & 1.631\\
S$_2$  & 0.499  & 0.54 & 0.853 & 0.972 & 0.977\\
Cl$_2$ & 0.262  & 0.30 & 0.446 & 0.531 & 0.544\\
\hline\hline
\end{tabular}
\end{flushleft}

\begin{flushleft}
$^a$ At CCSDT/cc-pVQZ bond distance, 1.2769 \AA, from Ref.\cite{Watts}.\end{flushleft}
\end{table}
\clearpage

\clearpage
\squeezetable
\begin{table}
\caption{Convergence of the contribution of valence connected quintuples ($T_5$) to the total atomization energy (kcal/mol), using various approximations\label{tab:T5}}
\begin{tabular}{l|ccc|ccc|ccc|ccc|cccc}
\hline\hline
& \multicolumn{3}{c}{CCSDTQ(5)$_\Lambda$$-$CCSDTQ} & \multicolumn{3}{c}{CCSDTQ(5)$-$CCSDTQ} & \multicolumn{3}{c}{$\hat{T_5}-(5)_\Lambda$}& \multicolumn{3}{c}{$\hat{T_5}-$(5)} & \multicolumn{4}{c}{$\hat{T_5}$ total}\\
& DZ & PVDZ & PVTZ & DZ & PVDZ & PVTZ & DZ & PVDZ & PVTZ & DZ & PVDZ & PVTZ & DZ & PVDZ &PVDZ & PVTZ \\
& ~~ & ~~~~ & ~~~~ & ~~ & ~~~~ & ~~~~ & ~~ & ~~~~ & ~~~~ & ~~ & ~~~~ & ~~~~ & ~~ & (no d)&~~~~ & ~~~~ \\
\hline
B$_2$ &0.057 &0.055 & 0.065 & 0.068 & 0.049 & 0.040 & 0.027 & 0.020 & 0.022  & 0.015 & 0.026 & 0.048 & 0.084 & 0.078 & 0.075 & 0.066 \\
C$_2$ &0.304 &0.338 & 0.350 & 0.470 & 0.465 & 0.399 &-0.031 &-0.018 &-0.012  &-0.196 &-0.146 &-0.061 & 0.274 & 0.236 & 0.320 & 0.338 \\
BN$^a$ & 0.214 & 0.191 & 0.231 & 0.100 & -0.127 & -0.174 & -0.035 & -0.037 & -0.040 & 0.078 & 0.280 & 0.355 & 0.178 & 0.177 & 0.154 & 0.181\\
N$_2$ &0.105 &0.113 & 0.110 & 0.117 & 0.125 & 0.106 & 0.003 & 0.001 & -0.002  &-0.009 &-0.011 & 0.003 & 0.108 & 0.113 & 0.114 & 0.109 \\
O$_2$ &0.066 &0.097 &       & 0.075 & 0.108 & 0.116 & 0.010 & 0.006 &        & 0.001 &-0.005 &       & 0.076 & 0.092 & 0.103 &  \\
F$_2$ &0.032 &0.039 &       & 0.038 & 0.044 & 0.074 & 0.000 & 0.001 &        &-0.006 &-0.004 &       & 0.032 & 0.025 & 0.040 &  \\
CO    &0.058 &0.040 &       & 0.059 & 0.019 &-0.006 &-0.013 &-0.008 &        &-0.014 & 0.013 &       & 0.046 & 0.034 & 0.032 &  \\
CN    &0.110 &0.118 &       & 0.156 & 0.144 & 0.111 & 0.017 & 0.000 &        &-0.029 &-0.026 &       & 0.127 & 0.130 & 0.118 &  \\
HF    &0.010 &0.003 & 0.007 & 0.011 & 0.002 & 0.005 & 0.000 & 0.000 &-0.001 & 0.000 & 0.000 & 0.001 & 0.010 & 0.001 & 0.002 &  0.006\\
H$_2$O&0.006 &0.005 &       & 0.007 & 0.008 & 0.009 & 0.000 & 0.000 &        & 0.000 & 0.000 &       & 0.006 & 0.004 & 0.008 &  \\
P$_2$ & 0.093 & 0.119 &       & 0.103 & 0.104 &       & 0.006 & 0.007 &        &        -0.004 & 0.022 &       & 0.099 & 0.100 & 0.126 &   \\
S$_2$ & 0.026 & 0.054 &       & 0.025 & 0.050 &       & 0.005 & 0.003 &        &        +0.006 & 0.007 &       & 0.031 & 0.031 & 0.057 &   \\
Cl$_2$ & 0.003 & 0.019 &       & 0.003 & 0.017  &       & 0.000 & 0.000 &        & 0.000 & 0.002 &       & 0.003 & 0.003 & 0.019 &   \\
\hline\hline
\end{tabular}
\begin{flushleft}
$^a$ At CCSDT/cc-pVQZ bond distance, 1.2769 \AA, from Ref.\cite{Watts}.
\end{flushleft}
\end{table}
\clearpage

\clearpage
\squeezetable
\begin{table}
\caption{Convergence of the differential contribution of quasiperturbative connected triple excitations, CCSD(T)$-$CCSD, to the core-valence component of the total atomization energy (kcal/mol)\label{tab:coreT}}
\begin{tabular}{l|ccccccccccc}
\hline\hline
& aug-pCVDZ & aug-pCVTZ & aug-pCVQZ & aug-pCV5Z & aug-pCV6Z & $\{$T,Q$\}$ & $\{$Q,5$\}$ & $\{$5,6$\}$ & SchwenkeTQ & SchwenkeQ5 & Schwenke56 \\
\hline
B$_2$  & 0.114 & 0.241 & 0.268 & 0.275 & 0.275 & 0.287 & 0.283 & 0.275 & 0.287 & 0.281 & 0.275 \\
C$_2$  & 0.347 & 0.642 & 0.698 & 0.712 & 0.712 & 0.738 & 0.728 & 0.712 & 0.738 & 0.724 & 0.712 \\
N$_2$  & 0.139 & 0.284 & 0.316 & 0.325 & 0.326 & 0.339 & 0.334 & 0.328 & 0.339 & 0.332 & 0.328 \\
O$_2$  & 0.091 & 0.185 & 0.206 & 0.212 & 0.213 & 0.222 & 0.218 & 0.215 & 0.222 & 0.217 & 0.215 \\
F$_2$  & 0.125 & 0.228 & 0.249 & 0.255 & 0.256 & 0.264 & 0.260 & 0.257 & 0.264 & 0.259 & 0.257 \\
CO     & 0.070 & 0.166 & 0.190 & 0.196 & 0.198 & 0.207 & 0.203 & 0.199 & 0.207 & 0.202 & 0.199 \\
CN     & 0.133 & 0.280 & 0.311 & 0.319 & 0.321 & 0.333 & 0.328 & 0.323 & 0.333 & 0.326 & 0.322 \\
NO     & 0.116 & 0.235 & 0.261 & 0.269 & 0.270 & 0.280 & 0.276 & 0.272 & 0.280 & 0.274 & 0.272 \\
HF     & 0.015 & 0.036 & 0.040 & 0.041 & 0.041 & 0.043 & 0.042 & 0.041 & 0.043 & 0.041 & 0.041 \\
H$_2$O & 0.028 & 0.059 & 0.066 & 0.067 & 0.068 & 0.071 & 0.069 & 0.068 & 0.071 & 0.069 & 0.068 \\
BH     & 0.032 & 0.053 & 0.058 & 0.060 & 0.060 & 0.062 & 0.061 & 0.060 & 0.062 & 0.061 & 0.060 \\
CH     & 0.027 & 0.047 & 0.051 & 0.052 & 0.052 & 0.054 & 0.053 & 0.052 & 0.054 & 0.053 & 0.052 \\
OH     & 0.019 & 0.039 & 0.043 & 0.044 & 0.044 & 0.046 & 0.045 & 0.045 & 0.046 & 0.045 & 0.044 \\
BF     & 0.016 & 0.049 & 0.059 & 0.062 & 0.063 & 0.067 & 0.066 & 0.065 & 0.067 & 0.065 & 0.064 \\
P$_2$  & 0.648 & 0.864 & 0.932 & 0.953 & N/A   & 0.982 & 0.976 & N/A   & 0.982 & 0.971 & N/A   \\
S$_2$  & 0.320 & 0.428 & 0.465 & 0.477 & N/A   & 0.492 & 0.490 & N/A   & 0.492 & 0.487 & N/A   \\
Cl$_2$ & 0.264 & 0.363 & 0.390 & 0.399 & N/A   & 0.409 & 0.408 & N/A   & 0.409 & 0.406 & N/A   \\
\hline\hline
\end{tabular}
\begin{flushleft}
\end{flushleft}
\end{table}

\clearpage
\squeezetable
\begin{table}
\caption{Higher order core-core and core-valence corrections (kcal/mol)\label{tab:coreHO}}
\begin{tabular}{l|cc}
\hline\hline
 & $\hat{T_3}-$(T) & $\Delta$(Q) \\
 & $\Delta$CV$\{$T,Q$\}$Z & CVTZ \\
 & (a) & (b)\\
\hline
B$_2$      & 0.035 & 0.072 \\
C$_2$      & 0.096 & 0.082 \\
N$_2$      & 0.021 & 0.013 \\
O$_2$      & 0.014 & 0.008 \\
F$_2$      & 0.012 & 0.007 \\
CO         & 0.020 & 0.018 \\
CN         & 0.026 & 0.033 \\
NO         & 0.017 & 0.017 \\
HF         &-0.001 & 0.005 \\
H$_2$O     & 0.002 & 0.005 \\
CH         & 0.006 & 0.000 \\
OH         & 0.001 & 0.003 \\
CH$_3$     & N/A   &-0.003 \\
CH$_4$     & N/A   &-0.004 \\
C$_2$H$_2$ & 0.022 & 0.009 \\
C$_2$H$_4$ & N/A   & 0.003 \\
NH$_3$     & N/A   & 0.002 \\
H$_2$CO    & N/A   & 0.012 \\
BN         & 0.088 & 0.116 \\
HNO        & N/A   & 0.015 \\
PH$_3$     & N/A   &-0.017 \\
Cl$_2$     & N/A   & 0.039 \\
ClF        & N/A   & 0.018 \\
HCl        & N/A   & 0.004 \\
S$_2$      & N/A   & 0.071 \\
CS         & N/A   & 0.084 \\
HS         & N/A   &-0.001 \\
H$_2$S     & N/A   &-0.001 \\
SO         & N/A   & 0.025 \\
\hline\hline
\end{tabular}
\begin{flushleft}
(a) ROHF reference. Values with UHF reference are very similar\\
(b) UHF reference.
\end{flushleft}
\end{table}

\clearpage
\squeezetable
\begin{table}
\caption{Comparison of W4.4 with other W4 variants and ATcT data for total atomization energies (kcal/mol)\label{tab:w4.4}}
\begin{tabular}{l|cccccccccc}
\hline\hline
& W4lite & W4 & W4.2 & W4.3 & $\Delta$[DBOC] & CV(Q) & W4.4$^a$ & W4.4$^b$ & ATcT & uncert. \\
&Ref.\cite{w4}&Ref.\cite{w4}&Ref.\cite{w4}&Ref.\cite{w4}&Ref.\cite{c2h6tae}&\multicolumn{3}{c}{Present work}&\multicolumn{2}{c}{Ref.\cite{w4}}\\
\hline
H$_2$     & 103.30 & 103.30 & 103.30 & 103.30 & -0.04 & 0.000 & 103.26 & 103.26 & 103.27 & 0.00 \\
OH        & 101.84 & 101.82 & 101.81 & 101.80 & -0.02 & 0.003 & 101.77 & 101.76 & 101.76 & 0.03 \\
H$_2$O    & 219.46 & 219.39 & 219.38 & 219.38 & -0.03 & 0.005 & 219.33 & 219.32 & 219.36 & 0.01 \\
C$_2$H$_2$& 388.57 & 388.72 & 388.72 & 388.79 & -0.03 & 0.009 & 388.73 & 388.70 & 388.62 & 0.07 \\
CH$_4$     & 392.52 & 392.52 & 392.52 & 392.53 & -0.04 &-0.004 & 392.47 & 392.45 & 392.50 & 0.03 \\
CH        & 80.01  & 80.02  & 80.02  & 80.03  & -0.02 & 0.000 & 80.00  & 79.99  & 79.98  & 0.05 \\
CO        & 256.17 & 256.19 & 256.18 & 256.21 & -0.01 & 0.018 & 256.17 & 256.15 & 256.25 & 0.03 \\
F$_2$     & 36.85  & 36.84  & 36.87  & 36.97  &  0.00 & 0.007 & 36.95  & 36.94  & 36.91  & 0.07 \\
HF        & 135.40 & 135.33 & 135.32 & 135.30 & -0.02 & 0.005 & 135.27 & 135.27 & 135.27 & 0.00 \\
N$_2$     & 224.90 & 225.01 & 225.00 & 225.07 & -0.01 & 0.013 & 225.02 & 224.99 & 224.94 & 0.01 \\ 
NH$_3$    & 276.62 & 276.60 & 276.59 & 276.61 & -0.04 & 0.002 & 276.55 & 276.53 & 276.59 & 0.01 \\
NO        & 149.74 & 149.81 & 149.81 & 149.86 & -0.01 & 0.017 & 149.83 & 149.80 & 149.82 & 0.02 \\
O$_2$     & 117.77 & 117.88 & 117.89 & 118.01 &  0.00 & 0.008 & 117.98 & 117.95 & 117.99 & 0.00 \\
Cl$_2$    & 56.85  & 57.03  & 57.01  & 57.08  &  0.00 & 0.039 & 57.08  & 57.07  & 57.18  & 0.00 \\ 
HCl       & 102.20 & 102.23 & 102.22 & 102.23 & -0.01 & 0.004 & 102.21 & 102.20 & 102.21 & 0.00 \\
H$_2$S    & 173.54 & 173.60 & 173.60 & 173.64 & -0.02 &-0.001 & 173.59 & 173.59 & 173.55 & 0.07 \\
SO        & 123.52 & 123.66 & 123.69 & 123.75 & -0.01 & 0.025 & 123.72 & 123.70 & 123.72 & 0.02 \\
C$_2$     & 143.88$^c$ & 143.86$^c$ & 144.03$^c$ & 144.08$^c$ &  0.00$^c$ & 0.082 & 144.08 & 144.07 & 144.03$^d$ & 0.13 \\

\hline\hline
\end{tabular}
\begin{flushleft}
(a) Using the usual partial-wave extrapolations for CCSD(5,6) and (T)(5,6).\\
(b) Using Schwenke's extrapolations for CCSD(5,6) and (T)(Q,5). Using (T)/(5,6) instead leaves results unchanged to two decimal places, except for CH$_4$, F$_2$, N$_2$, and O$_2$ (+0.01 kcal/mol each) and H$_2$S and C$_2$ (-0.01 kcal/mol each).\\
(c) Present work.\\
(d) B. Ruscic, personal communication quoted in Ref.\cite{Feller2007}.\\

When using CCSD(T)/cc-pwCVQZ reference geometries (all electrons correlated except the $1s$ deep-core orbitals on second-row atoms), dissociation energies at all levels are found to go up by 0.03 kcal/mol for Cl$_2$ and SO, by 0.02 kcal/mol for C$_2$H$_2$, CO, and N$_2$, by 0.01 kcal/mol for five additional molecules (namely, CH$_4$, NH$_3$, NO, O$_2$, and ClF).
\end{flushleft}
\end{table}


\begin{thebibliography}{99}
\bibitem{carbon-av7z} Note that the published carbon AV7Z basis set is a local minimum and will exhibit anomalous convergence behavior in extrapolations. A revised basis set was communicated to the present authors by David Feller and is in essentially complete agreement with our own independent reoptimization. The revised basis set will be made available online at the EMSL Basis Set Exchange.
\end{thebibliography}
\end{document}